# Luminescence evidence for bulk and surface excitons in free xenon clusters

O.G. Danylchenko, Yu.S. Doronin, S.I. Kovalenko,

M.Yu. Libin, V.N. Samovarov[*], V.L. Vakula

*B. Verkin Institute for Low Temperature Physics and Engineering*

*of the National Academy of Sciences of Ukraine,*

*47 Lenin Ave., Kharkiv, 61103, Ukraine*

## Abstract

Cathodoluminescence spectra of free xenon clusters produced by condensation of xenon-argon gas mixtures in supersonic jets expanding into vacuum were studied. By varying initial experimental parameters, including xenon concentration, we could obtain clusters with a xenon core (300-3500 atoms) covered by an argon outer shell as well as shell-free xenon clusters ($\approx$1500 atoms). The cluster size and temperature ($\approx$40 K for both cases) were measured electronographically. Luminescence bands evidencing the existence of bulk and surface excitons were detected for shell-free xenon clusters. The emission from bulk excitons in small clusters is supposed to be due to processes of their multiple elastic reflections from the xenon-vacuum interface. A presence of an argon shell causes extinction of the excitonic bands. In addition, some new bands were found which have no analogs for bulk xenon cryosamples.

---

[*] Corresponding author: samovarov@ilt.kharkov.ua

# I. INTRODUCTION

In bulk cryocrystals of heavy rare gases like argon, krypton, and xenon, excitonic states have been observed both in absorption and luminescence spectra [1]. Structure defects, surface, impurities make it much more difficult to register excitonic features in luminescence spectra, which suggests strong damping of free excitons on such lattice imperfections. In clusters of these gases, the excitonic states of both bulk and surface type have been observed so far only in absorption spectra [2,3], e.g., a dependence of energy position of n>1 exciton levels on cluster radius (confinement effect) was reported in Ref. [3].

The motivation of this paper grew out of the electron-diffraction and cathodoluminescence results of our previous works [4,5] and is to detect excitonic luminescence in free xenon clusters. In Ref. [4] it was shown that condensation of argon-xenon mixtures into clusters can be accompanied by a complete phase segregation into a xenon cluster core and an argon outer shell with a sharp interface between the components. At the same time, Ref. [5] demonstrated that an increase in heavy-impurity concentration in the primary gas mixture can cause formation of pure krypton clusters of a lowered temperature, argon serving only as a carrying gas. Similar results obtained earlier using X-ray photoelectron spectroscopy were reported in Ref. [6] for a phase segregation in argon-xenon clusters and Ref. [7] for elimination of argon clusters in argon-krypton mixtures.

Temperature is an important factor in observation of excitonic bands in luminescence spectra, since cooling intensifies strongly the excitonic luminescence in bulk xenon below $T \approx 60$ K [8]. Using argon-xenon mixtures with the pressure ≥1 atm and the initial xenon concentration over 2 at.% allowed us to obtain pure xenon clusters with a temperature lowered down to about 40 K, as well as xenon clusters of the same temperature, but covered by a thin argon shell. It was in these experiments that we detected for the first time luminescence bands being due to bulk and surface excitons in xenon. Some of the results obtained were reported briefly in Ref. [9]. Here we present a more complete experimental picture with an extended analysis of the observed features. We believe that our results can help to open new opportunities in the studies of excitonic states, including those of exciton-related quantum size effects, in substrate-free rare-gas clusters.

## II. EXPERIMENT

Experiments were carried out on free clusters obtained by the method of isentropic expansion of a supersonic gas jet (see, for example, Ref. [10]). A beam of a mixture of argon and xenon gases exhausted into vacuum through a metallic conical nozzle with the throat diameter 0.34 mm, cone angle 8.6°, and exit-to-throat cross-sectional area ratio 36.7. The vacuum chamber was evacuated with a condensation pump cooled by liquid hydrogen. The pressure $p_0$ and temperature $T_0$ of the gas mixture at the entrance to the nozzle was varied from 0.5 to 2.5 atm and from 160 to 250 K, respectively. The concentration of xenon in a mixture $C_{Xe}$ was varied from 1 to 3 at.%. Cluster beams were probed at the distance of 30 mm from the nozzle exit. Depending on the initial parameters $C_{Xe}$, $T_0$, and $p_0$, it was possible to obtain xenon clusters with practically no argon and those covered by an argon shell. Increase in pressure and/or decrease in temperature of the gas mixture favored formation of larger clusters.

### A. Electron-diffraction spectroscopy

Electron diffraction measurements allowed us to observe diffraction patterns of clusters up to the diffraction vector values $s \approx 7$ Å$^{-1}$, $|s| = \lambda^{-1} \cdot 4\pi \sin\nu$, where $\nu$ is diffraction angle, $\lambda$ is electron wavelength (electron energy was 50 keV). In the case of mixed argon-xenon clusters, the relative quantity of the chemical components was estimated from the ratio of argon and xenon diffraction peak intensities. The difference between the atomic factors of electron scattering by these substances was taken into account. The cluster size was estimated by broadening of the diffraction peaks. Electron-diffraction measurements also made it possible to determine the cluster temperature $T_{cl}$. With this purpose we measured lattice parameters of argon or xenon to compare them with the values derived from the relevant well-known temperature dependences. The background resulting from incoherent (nonelastic) and gas scattering of electrons was subtracted from the diffraction patterns.

### B. VUV cathodoluminescence spectroscopy

Clusters were optically studied by their cathodoluminescence spectra measured in the region 8-8.5 eV, where the atomic emission lines of Xe ($^3P_1$) and emission bands from bulk and surface excitons in massive xenon samples are located. We also measured luminescence spectra in the vicinity of 7 eV, where the molecular $Xe_2^*$ emission band is observed. All measurements were made for constant values of electron energy (1 keV) and beam current (20 mA), which allowed us to make quantitative comparisons between the spectra obtained under different initial conditions. In this paper we pay main attention to cathodoluminescence measurements performed for xenon concentrations in the primary gas mixture $C_{Xe} \geq 1\%$. The data for smaller concentrations were reported in our paper [4] devoted to the problem of phase segregation in argon-xenon clusters resulting in formation of a pure xenon core.

### III. RESULTS

By varying the initial experimental parameters (such as xenon concentration in the primary argon-xenon gas mixture, temperature and pressure of the mixture at the nozzle entrance), we were able to obtain xenon clusters of two types: those covered by an argon shell and those with no argon. The shell-free clusters were obtained only in a certain range of xenon concentration, while the clusters with an argon shell emerged on deviating towards higher or lower values. The size of the xenon core in the phase-separated argon-xenon clusters was a function of gas parameters at the nozzle entrance and could differ by an order of magnitude, giving rise to noticeable distinctions in the luminescence spectra. Based on composition of the observed spectra, we divided them into three categories characterized by dynamics of electronic excitations: shell-free xenon clusters, small-, and large-sized xenon clusters with an argon shell.

### A. Small-sized Xe clusters with an Ar shell

For gas-mixture parameters at the nozzle entrance $C_{Xe}=1\%$, $T_0 \leq 245$ K, $p_0 \geq 1$ atm, clusters consist of two components: a xenon core and an argon shell. It should be noted that according to the electron-diffraction data and VUV luminescence spectra [4,11], the argon shell does not contain any xenon impurity, while the xenon core is practically argon-free, i.e. there occurs a phase segregation into two pure components with a sharp interface between them. This is evidenced, in particular, by the fact that there is no emission of the heteronuclear molecule $(ArXe)^*$ and of the Xe molecules in the Ar matrix $(Xe_2^*)_{Ar}$.

It should also be noted that, as is shown in Refs. [4,5], in the case of condensation of binary Ar-Kr and Ar-Xe mixtures the concentration of the heavier impurity in clusters exceeds that in the primary gas mixture by approximately one order of magnitude, i.e. for our case the xenon concentration in the clusters was about 10%.

Figure 1(a) shows electron diffraction patterns for $C_{Xe}=1\%$, $T_0=245$ K, $p_0=1.5$ and 2.5 atm. It is clearly seen that the peak at $s \approx 3.2$ Å$^{-1}$ is abnormally large with respect to the argon peak (311) at $s \approx 3.9$ Å$^{-1}$ (in pure argon the intensities of the two peaks should be similar). This is due to the fact that the enhanced intensity of the peak at $s \approx 3.2$ Å$^{-1}$ results from an overlapping of the argon peak (220) and the xenon peak (311). The argon and xenon (111) peaks merge into a broad maximum at $s \approx 2$ Å$^{-1}$. As the pressure is increased, the maximum at $s \approx 2.9$ Å$^{-1}$, assigned to the (220) xenon peak, gets more pronounced [12]. So even at elevated temperatures of the primary gas mixture the resulted clusters are mixed. As the temperature is lowered, the xenon core peaks intensify and narrow, i.e. the volume of the xenon core increases. According to our electronographic estimates, at $T_0 \approx 160$ K the number of atoms in the xenon core is about several hundreds (approximately 4 Mackay spheres, $r_{core} \approx 16$ Å), while the number of atoms in the argon shell is 2000-3000. The cluster temperature was estimated as $T_{cl} \approx 40$ K from the measured lattice parameter of argon.

Figure 1(b) shows the cathodoluminescence spectra for the relevant xenon concentration at $T_0=170$ K, $p_0=1.5$ and 2 atm. The three observed spectral features are centered at 8.268 eV (Gaussian dispersion is 0.035 eV), 8.41, and 8.44 eV. As the pressure is increased, the 8.27 eV component gains more intensity, and so does the 8.41 eV component with respect to the 8.44 eV one. The 8.44 eV band corresponds to the emission of $Xe^*$ ($^3P_1$) atoms desorbed from the cluster. The 8.41 eV component, located at the redder side from the desorption line, should be assigned to the emission of $Xe^*$ ($^3P_1$) centers in the outer minimum near the argon surface. Earlier, similar emission centers of $Kr^*$ ($^3P_1$) were observed near the argon surface for bulk cryocrystals grown from argon-krypton mixtures [13]. It is difficult to give an unequivocal assignment of the 8.27 eV band to a certain emission center, however, as our data demonstrate (see also below), the feature arises as an argon shell appears.

### B. Shell-free Xe clusters

An increase in xenon concentration in the primary gas mixture gives rise to the situation that argon does not stick to xenon clusters due to their substantial heating resulting from significant xenon condensation energy emission. The argon serves here only as a carrier gas removing heat from clusters. A similar regime of cluster formation was earlier observed also for argon-krypton mixtures upon an increase in krypton concentration [5,7].

Figure 2(a) shows the electron diffraction pattern for $C_{Xe}$=3%, $T_0$=165 K, and $p_0$=1 atm. The pattern reveals no argon peaks, while the xenon peaks are well pronounced. The number of atoms of xenon clusters was estimated to be $N\approx1500$ ($r_{cl}\approx27$ Å). The xenon cluster temperature, as judged from the lattice parameter estimates, was, similarly to the previous case, about 40 K.

Let us start our consideration of pure xenon clusters from the cases of $T_0$=175 K, $p_0$=0.75 atm, and $T_0$=180 K, $p_0$=0.5 atm. The results are shown in Fig. 2(b) together with their decomposition into Gaussian components. There are no mentioned above features at 8.27 eV and 8.41 eV in the spectra, but there still remains a rather strong 8.44 eV component of the emission of desorbed atoms. Additionally, 4 new components appear at: 8.245 eV (Gaussian dispersion is 0.033 eV), 8.296 eV (0.03 eV), 8.315 eV (0.018 eV), and 8.352 eV (0.016 eV). Thus, the emission spectrum of pure xenon clusters turns out to be more complicated. It can be seen that an increase in pressure at a virtually constant temperature (i.e., an increase in cluster size) causes the most pronounced change in the 8.35 eV component, whose intensity grows by a factor of 2.2.

Figure 2(c) shows the spectra at the constant pressure of 1 atm, but for temperatures lowering from 225 K down to 170 K. All the 5 components described above for pure xenon clusters are clearly seen. The substantial lowering of temperature favors an increase in cluster size. The most pronounced change was observed for the 8.296 eV component, whose intensity grew by almost 6 times, and for the 8.35 eV component, which narrowed from its Gaussian dispersion value 0.023 eV ($T_0$=225 K) to 0.017 ($T_0$=170 K). So, for both temperature and pressure variations within the above ranges all the components persist, but the intensity ratio between them changes.

Based on the data of VUV spectroscopy for bulk xenon cryocrystals (see, for example, Ref. [1]), we can rather unequivocally assign the 8.35 eV band to the emission of a free $n$=1 exciton $\Gamma(3/2)$. The two other bands at 8.245 and 8.315 eV can be due to the $n$=1 surface exciton splitting in the crystal field in accordance with Ref. [14]. It should be noted that for pure xenon clusters, in consistence with our expectations, the 8.27 eV and 8.41 eV bands, arising from an argon shell, are no more observable. The nature of the 8.3 eV band, which was not detected earlier in optical spectra of bulk xenon, remains unclear.

### C. Large-sized Xe clusters with an Ar shell

Increase in initial gas pressure $p_0$ with other experimental parameters unchanged results in formation of an argon shell around the xenon core. Figure 3(a) shows the electron-diffraction pattern for $C_{Xe}$=3%, $T_0$=160 K, and $p_0$=2 atm. The argon (200) peak is clearly seen, the xenon (200) and (311) peaks are abnormally intensified due to their overlapping with the argon (111) and (220) peaks, respectively. Taken as a whole, the pattern provides evidence of a xenon core covered by crystalline argon. Electron diffraction data allow us to estimate the core as consisting of $N \approx 3500$ atoms ($r_{core} \approx 36$ Å) and the number of atoms in the argon shell as 7000-8000. The cluster temperature was about 40 K as in the previous cases.

Figure 3(b) shows the cathodoluminescence spectra for $T_0$=170 K, $p_0$=2 atm, and $T_0$=175 K, $p_0$=1.25 atm. For the case of $p_0$=1.25 atm there is a band at 8.265 eV (Gaussian dispersion is 0.017 eV), which dominated the spectrum of the phase-separated argon-xenon clusters formed in gas mixtures with $C_{Xe}$=1%. Some traces of the line at 8.41 eV can be seen in the spectrum suggesting some emission from Xe$^*$ ($^3P_1$) atoms in the outer minimum near the argon surface. These two features reflect the fact that at this pressure xenon clusters start getting covered with an argon shell. The excitonic line at 8.35 eV becomes twice as weak as that at $p_0$=1 atm (see Fig. 2(c)).

A practically complete extinction of the excitonic band and a significant growth of the surface line at 8.41 eV as well as of the band at 8.265 eV are observed for $p_0$=2 atm (see Fig. 3(b)). Nevertheless, even in the presence of an argon shell, the 8.317 eV band and undoubted traces of the 8.23 eV band are detected in the spectrum. Hence surface excitons (8.317 and 8.23 eV) are still present in a xenon core covered by an argon shell, while free bulk excitons are significantly suppressed.

## IV. DISCUSSION

For the main quantum state n=1 the radius of excitons in argon, krypton, and xenon is smaller than the nearest neighbor distance $d$ and is $r^{n=1}$ =3.2 Å for xenon ($d_{Xe}$=4.34 Å). Thus the exciton radius turns out to be smaller than the size of xenon cores/clusters in all our measurements. The lifetime of the n=1 exciton before it self-traps into molecular states like those of $Xe_2^*$ is about $10^{-9}$ s, and the distance it goes during its lifetime is several hundreds of ångströms [14], which exceeds substantially the xenon core/cluster linear parameters measured in our experiments. Thus the probability of emission of a free exciton in our experiments will be determined mainly by mechanisms of its interaction with the xenon-vacuum or xenon-argon interface.

For the case of a small xenon core covered by an argon shell ($C_{Xe}$=1% and $T_0 \leq$245 K in our experiments), we observed neither surface nor bulk excitons, which can be explained by the small size of the core, its icosahedral structure, and a strong exciton damping effect of the argon shell. However, even under these conditions we detected emission of desorbed and above-surface atoms $Xe^*$ ($^3P_1$), whose observation wasn't reported earlier for bulk xenon. As we pointed out earlier, in small-sized Ar-Xe clusters the argon shell contains virtually no xenon, therefore it can be thought that an excited atom of xenon, possessing a kinetic energy, penetrates through the argon shell. This effect can be favored by the negative electron affinity in the excited atom to the argon matrix [13]. A similar behavior was observed for Ar* atoms that could move through the Ne cluster, desorb and emit light in the vacuum [15].

Well-pronounced emission bands from surface and bulk excitons are observed only for shell-free fcc-structured Xe clusters having a temperature of about 40 K, when there is no damping effect of an argon shell. Taking into account that the size of a xenon cluster was about 30 Å (more than an order of magnitude smaller that the average distance an exciton can move from point of generation until it is trapped), a moving exciton should undergo multiple elastic reflections from the vacuum interface before it emits a photon inside the cluster. The halfwidth of the excitonic line at 8.35 eV was ≈0.02 eV in our measurements, which is twice as big as that for bulk samples (≈0.01 eV) [8]. This broadening is likely due to additional structure defects in a xenon cluster with respect to perfect bulk crystals.

Simultaneously with the appearance of the 8.35 eV excitonic line, we observed a decrease in intensity of the emission from $Xe_2^*$ molecules at 7.1 eV (which is also typical of bulk xenon samples [16]). This fact suggests that an argon shell favors trapping of excitons into quasimolecular states, while its absence reinforces the channel of their elastic reflection from the surface. In other words, an argon shell lowers sharply the barrier of exciton localization into quasimolecular states.

The same conclusion can be made from our measurements on large-sized xenon clusters covered by an argon shell. Although the size of the xenon core is rather great in this case (about 40 Å) and crystalline, it is still the argon shell that is responsible for the extinction of the bulk exciton line at 8.35 eV. The surface excitons peaked at 8.23 and 8.32 eV are significantly less influenced by the argon shell than the bulk excitations.

A clear assignment of the emission bands centered at 8.27 and 8.3 eV is hard to be made so far. We can only affirm that the 8.27 eV band emerges only in the presence of an argon shell and may be due to some interface excitations. Earlier, the interface excitations in absorption experiments were observed in Ref. [17] for neon-covered argon clusters prepared by a pick-up technique. On the other hand, it is possible that one of these bands is due to the forbidden transition from $Xe^*$ ($^3P_2$) state (located at 8.315 eV in free atoms), which is allowed by symmetry breaking near the surface.

## V. CONCLUSIONS

We observed for the first time luminescence response from bulk (8.35 eV) and surface (8.23 and 8.31 eV) n=1 excitons in free xenon clusters. In small clusters with $r_{cl} \approx 30$ Å, the bulk excitons undergo multiple elastic reflections from the xenon-vacuum interface before they emit photons inside the cluster. The effect of multiple elastic reflections favoring the direct observation of free excitons seems to be rather unusual, since the surface of bulk cryocrystals efficiently suppresses bulk excitons. The surface excitons we detect in our experiments are less sensitive to the argon shell, their weak emission bands are also observable in the case of a xenon core covered by a thin argon shell, when bulk excitons are strongly suppressed. We also found that the 8.27 eV luminescence band is rather sensitive to an argon shell and can serve as a probe to study phase segregation in mixed clusters. Another effect that we observe lies in the fact that excited atoms $Xe^*$ ($^3P_1$) possessing large kinetic energy can penetrate through the argon shell and emit photons outside the cluster either as desorbed atoms (at 8.44 eV), or as atoms located in the outer minimum above the argon surface (at 8.41 eV).

## ACKNOWLEDGMENTS

We gratefully acknowledge fruitful discussions with E.V. Savchenko, E.I. Tarasova, and A.G. Belov.

**Figure captions**

FIG. 1. (*a*) Electron diffraction patterns from small phase-separated argon-xenon clusters produced in a supersonic jet with the xenon concentration $C_{Xe}$=1% in the primary gas mixture with the temperature $T_0$=245 K at the nozzle entrance. Arrows point to the diffraction peaks. The pressure at the nozzle entrance was $p_0$=1.5 and 2.5 atm. (*b*) Cathodoluminescence spectra of small-sized xenon clusters covered by an argon shell for the initial parameters $C_{Xe}$=1%, $T_0$=170 K, $p_0$=1.5 and 2 atm. Thin solid lines are Gaussian fits to the data (8.268 eV component).

FIG. 2. (a) Electron diffraction pattern from shell-free xenon clusters for the primary mixture parameters $C_{Xe}$=3%, $T_0$=165 K, and $p_0$=1 atm. The number of xenon atoms in clusters was about 1500. (b) Cathodoluminescence spectra of shell-free xenon clusters were obtained for the gas mixture parameters: $C_{Xe}$=3%, $T_0$=180 K, $p_0$=0.5 atm, and $C_{Xe}$=3%, $T_0$=175 K, $p_0$=0.75 atm. (c) Temperature change in the cathodoluminescence spectra at pressure $p_0$=1 atm and xenon concentration $C_{Xe}$=3% for $T_0$=225 K and 170 K. Thin solid lines are Gaussian fits to the data.

FIG. 3. (*a*) Electron diffraction pattern from large-sized xenon clusters with an argon shell (the xenon core contains ≈3500 atoms). Gas mixture parameters at the nozzle entrance: $C_{Xe}$=3%, $T_0$=160 K, and $p_0$=2 atm. (b) Cathodoluminescence spectra of clusters getting covered by an argon shell with growing pressure: $C_{Xe}$=3%, $T_0$=175 K, $p_0$=1.25 atm, and $C_{Xe}$=3%, $T_0$=170 K, $p_0$=2 atm. Thin solid lines are Gaussian fits to the data.

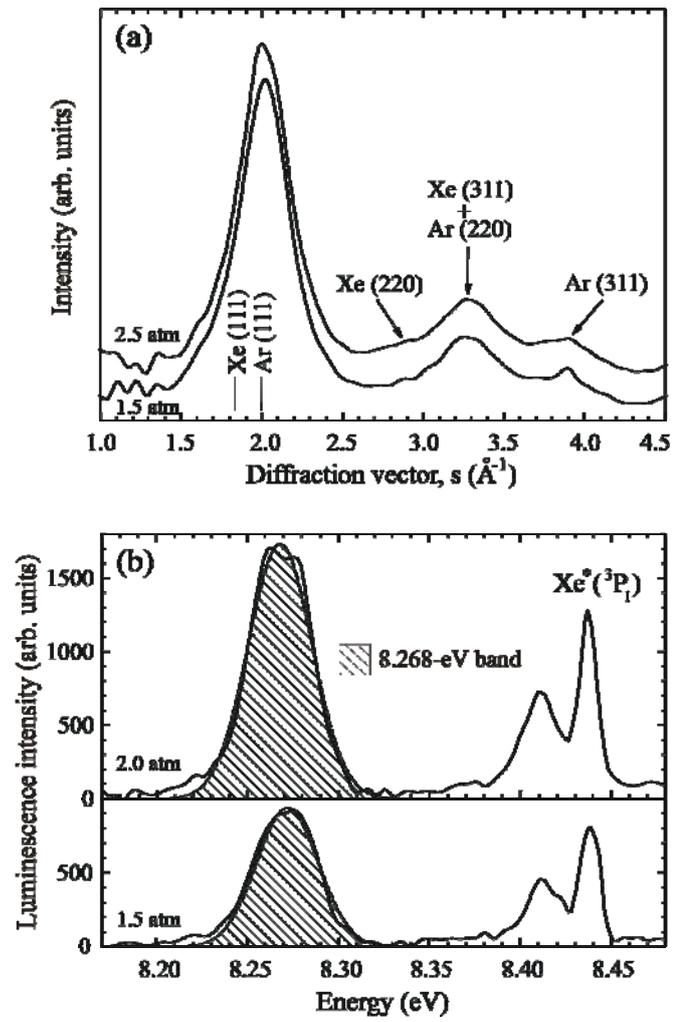

Fig. 1

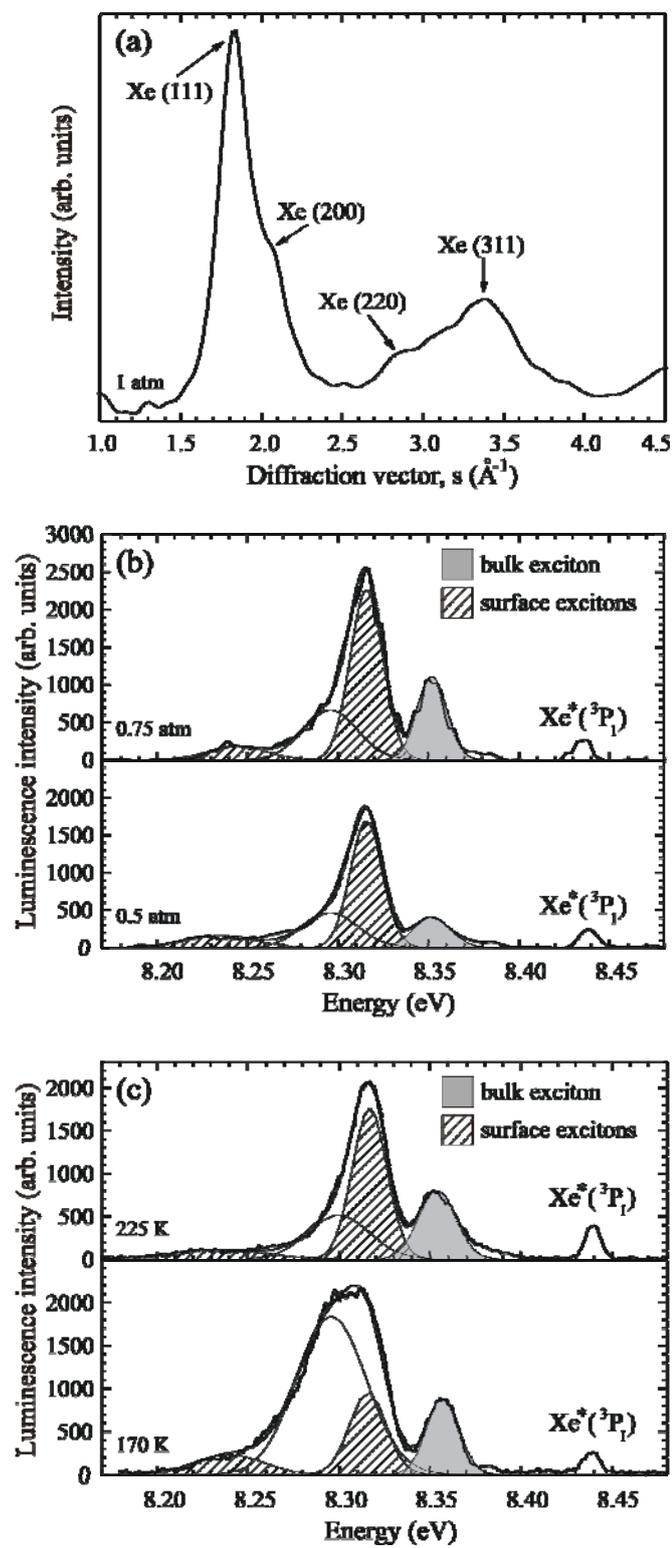

Fig. 2

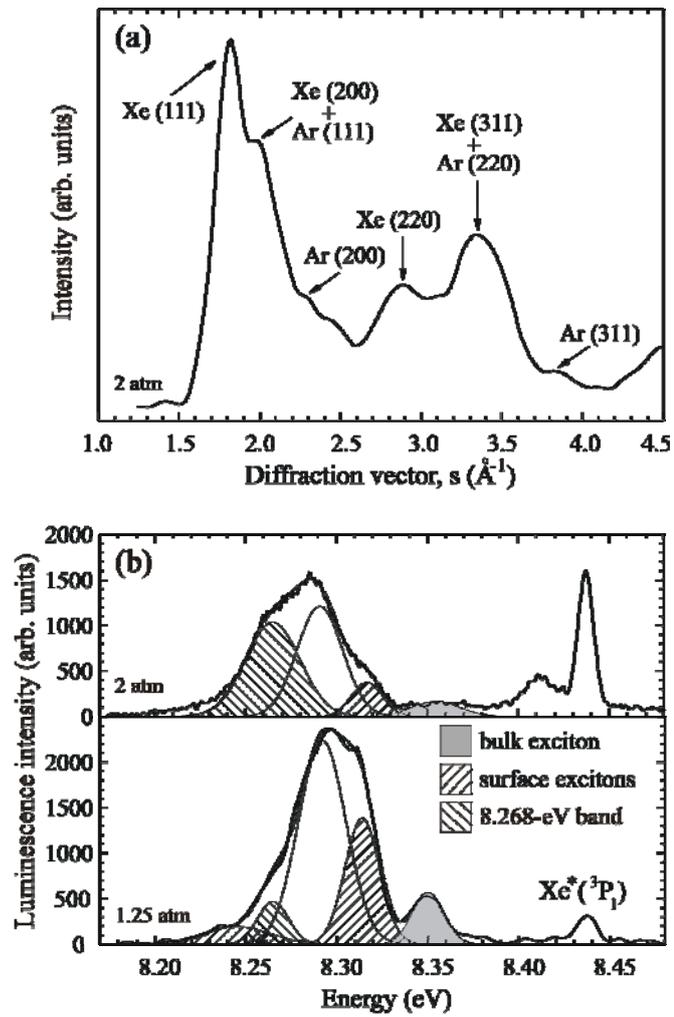

Fig. 3